ARTIFICIAL INTELLIGENCE

# The science and practice of proportionality in AI risk evaluations

AI evaluations should provide meaningful risk information without imposing excessive burden

Carlos Mougan[1], Lauritz Morlock[1], Jair Aguirre[2], James R. M. Black[3], Jan Brauner[1], Simeon Campos[4], Sunishchal Dev[2], David Fernández Llorca[5], Alberto Franzin[1], Mario Fritz[6], Emilia Gómez[5], Friederike Grosse-Holz[1], Eloise Hamilton[1], Max Hasin[1,7], Jose Hernandez-Orallo[8,9], Dan Lahav[10], Luca Massarelli[1,11], Vasilios Mavroudis[12], Malcolm Murray[4], Patricia Paskov[2], Jaime Raldua[13], Wout Schellaert[1]

A global challenge in artificial intelligence (AI) regulation lies in achieving effective risk management without compromising innovation and technical progress (*1*). The European Union (EU) Artificial Intelligence Act (*2*) represents the first regulatory attempt worldwide to navigate this tension in the form of a binding, risk-based framework. In August 2025, obligations for providers of general-purpose AI (GPAI) models under the EU AI Act entered into application. They require providers of the most advanced GPAI models to evaluate possible systemic risks stemming from their models (*3*). This raises the regulatory challenge of ensuring that the evaluations provide meaningful risk information without imposing excessive burden on providers. The principle of proportionality, a binding requirement under EU law, requires the regulator to calibrate its actions to their intended objectives. The application of proportionality to model evaluations for AI risk opens opportunities to develop scientific methods that operationalize such calibration within concrete evaluation practices.

According to the principle of proportionality, EU measures must be suitable, necessary, and balanced. A measure is suitable if it pursues an objective that is aligned with the intent of its legislative basis and is implemented consistently and systematically; in other words, it must meet a minimum level of effectiveness for a legitimate aim. It is considered necessary for the level of effectiveness that it achieves if there is no less restrictive yet equally effective means to achieve its objective; in other words, it cannot be a case of a sledgehammer cracking a nut. It is balanced if its burden does not clearly outweigh its benefit; in other words, even if a measure is suitable and necessary, it must not appear excessive (*4*). Regulators have some discretion in determining whether a measure meets these conditions, if doing so requires a complex and evidence-based case-by-case assessment that a court would not be better placed to replicate in hindsight (*5*).

Recent work [e.g., (*6*)] emphasizes the importance of evidence-based AI policy and the corresponding need to accelerate evidence generation. Nevertheless, it does not address the regulatory question of how much evaluation effort is legally justifiable, i.e., of what proportionality means in practice. Although this question arises in other regulatory domains as well, applying it to AI model evaluations poses distinct challenges (*7*). In the absence of methodologies to determine when requiring an evaluation is proportionate, regulators face limited guidance in navigating complex trade-offs, expanding the scope for normative judgment and increasing the risk of regulatory outcomes that are insufficiently protective or unduly burdensome.

## PROPORTIONALITY FOR AI RISK EVALUATIONS

The AI Office of the European Commission convened a series of workshops to examine how the principle of proportionality can be applied to model evaluations for AI risk in practice. The resulting findings point to a gap at the intersection of law, science, and evaluation practice: the absence of clear and scientifically grounded methodologies to determine when requiring an evaluation is proportionate.

### Suitability

A suitable evaluation contributes consistently and systematically to the assessment of a risk posed by the model, reaching a minimum level of effectiveness (see the figure). Effectiveness can be measured through informational value, i.e., the information that an evaluation provides regarding the risk that it is intended to assess within the model. Because risk is typically shaped by the AI model and the context in which it is developed and deployed, such as its access conditions or platform integration, an evaluation must take these factors into account to provide meaningful information about the risk. Moreover, because the model's outputs may lead to a range of downstream effects, only some of which are potentially harmful (*8*), an evaluation must consider the risk pathway from the model to harm to provide meaningful information. In practice, the following four criteria can be considered for assessing whether an evaluation provides sufficient informational value to warrant its suitability for a given model (*9*).

**Realistic** The evaluation is designed to reflect real-world conditions and operational constraints. For example, for the risk of real-world cyber operations, the evaluation takes into account noisy inputs, system configurations, incomplete information, or the human factor.

**Sensitive** The evaluation can detect meaningful changes in the model's performance. For example, evaluations that are already saturated, or so difficult that they make it hard to measure progress, yield limited informational value compared to those that can reliably capture incremental improvements.

**Specific** The evaluation is sufficiently tied to the risk under assessment, for example, by evaluating model capabilities that are relevant to an attack chain or risk scenario.

**Rigorous** The evaluation is conducted in accordance with measurement science. This implies ensuring the evaluation's validity, reliability, and alignment with technical standards.

Given that these criteria are not binary, the extent to which each needs to be met for the evaluation to be deemed suitable is to be determined on a case-by-case basis, with input from the scientific community.

### Necessity

Assessing necessity requires comparing the burden of carrying out the evaluation to that of carrying out evaluations with equal or greater effectiveness for assessing the risk in question (inter-evaluation comparison). For a given level of effectiveness, an evaluation is necessary if there is no other evaluation that is at least as effective and imposes a lower burden (see the figure). The following two considerations are relevant for assessing and comparing the burden of evaluations.

**Level of intrusiveness** The extent to which an evaluation interferes with the provider's business, the model infrastructure, or users' experience needs to be taken into consideration. Examples of such interference could include the need to access sensitive model components or disrupt model deployment to carry out the evaluation.

**Resources** Both the resources required to develop an evaluation and those needed to execute it up to the desired effectiveness level need to be considered. Development costs relate to the effort necessary to design, implement, and integrate an evaluation within a provider's existing infrastructure. For example, this could include developing evaluation methodologies or datasets. Execution costs concern the resources required to run the evaluation once developed, including compute, time, and human involvement.

Assessing and comparing effectiveness is challenging because the informational value provided by an evaluation for a given model is multidimensional. As a result, an evaluation may be more informative along certain risk dimensions while being less informative or less reliable along others. Moreover, evaluations are empirical measurements whose results are subject to uncertainty and can be affected by implementation choices such as sampling strategies or protocol design. For these reasons, comparing the informational value of different evaluations cannot rely on a single metric or binary notion of effectiveness. Rather, it first requires a methodological approach that identifies a coherent set of metrics, each capturing distinct dimensions of informational value, and that takes into account measurement uncertainties. Second, where empirical evidence remains incomplete or ambiguous, it requires a normative judgment to determine whether the remaining differences in informational value are relevant for the specific risk, model, and context under consideration. If the differences in informational value are negligible, then evaluations could be considered equally effective, with only the least burdensome one thus being considered necessary.

Research is required to enable the comparison of informational value and burden across evaluations, and in doing so to provide methodologies for establishing when an evaluation is necessary.

### Balancing

The informational value of a specific evaluation must be weighed against its burden to assess whether there is a manifest imbalance (intra-evaluation comparison). If so, another evaluation with a lower effectiveness level and lower burden may be more appropriate (see the figure). The following considerations can guide the weighing of informational value against burden for an evaluation.

**Risk profile of the model** The estimate of the probability and severity of harm stemming from the model may change throughout the model's life cycle and risk assessment process, based on new information. Similarly, the level of confidence in the risk estimates may change throughout the risk assessment cycle. The evaluation should reflect the most up-to-date risk estimate of the model, taking into account the level of confidence in that estimate. In the absence of prior risk estimates, early risk indicators such as those based on the model's training data, volume, development techniques, or the nature and scale of deployment (such as open-source availability, system integrations, or reach in number of potential users) could serve as a basis for choosing a specific evaluation.

**Provider capacity and relative burden** Burden is not absolute, but should rather be weighed relative to the size and capacity of the affected provider, to the release and distribution strategy of the model, and to the extent to which constraints are self-inflicted.

On the basis of these considerations, a tiered evaluation methodology can help avoid misjudgments. The evaluation process could begin with low-cost and minimally intrusive methods (such as static prompting or simple multiple-choice tests) for which failure may be informative, e.g., sufficient to rule out elevated risk. If a model convincingly and unambiguously fails in simplified tasks, it is unlikely to succeed in real-world

## Proportionality in risk evaluations

The burden and effectiveness of an evaluation are dependent on a number of factors, e.g., scaffolding techniques used or sample size. Each curve represents an evaluation method, and each point on the curve is an evaluation. A suitable evaluation must exceed the minimum level of effectiveness. An evaluation is necessary for a given effectiveness level if it is on the Pareto frontier of burden–effectiveness. The balancing step is to select a point on this frontier, an evaluation that reaches the level of effectiveness deemed most appropriate given the burden it incurs.

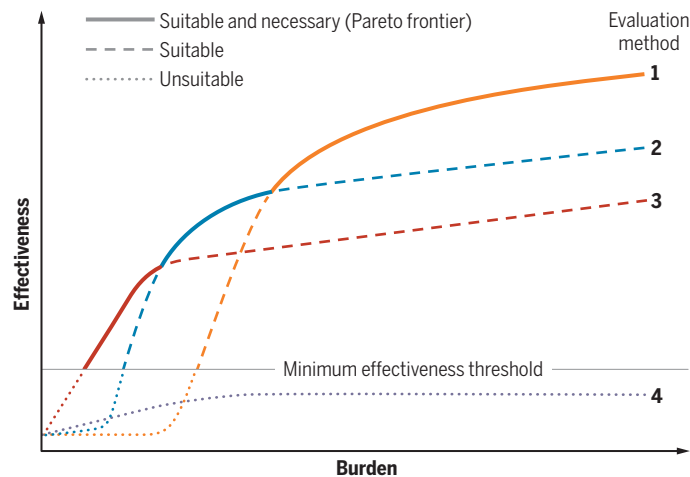

conditions. When initial evaluations prove inconclusive, reveal saturation, or suggest higher risk levels, more demanding methods become appropriate. These might include evaluations on more realistic tasks with extended run-times or tool access. By progressively increasing the evaluation burden, this iterative process continues until evaluators achieve sufficient confidence in their risk estimate (*10*). This iterative approach also aligns with the precautionary principle, a binding administrative principle in EU law. To reconcile proportionality with precaution, continuing evaluations until reaching adequate confidence in the risk estimates carries considerable weight when balancing competing considerations (*11*).

Determining the level of informational value that an evaluation provides and weighing it against its burden is an open challenge. Advancing the science behind the balancing step requires further research, including the development of metrics and methodologies that can guide this weighing. Additionally, further research on how the resources or intrusiveness of evaluations can be lowered while maintaining their informational value will allow their broader balanced application.

### CASE STUDY: CYBER VULNERABILITY DISCOVERY

While GPAI models can present risk in a multitude of domains and contexts, the principle of proportionality encourages the identification of risks and scenarios in key specific cases that are representative and informative of this much wider range of situations. The simplified case study below is intended only as an example to clarify how proportionality can be applied to a specific risk scenario: a malicious actor using an AI model to identify and exploit previously unknown vulnerabilities in an open-source codebase.

### Suitability

We consider three distinct evaluation methods—HonestCyberEval (*12*), BountyBench (*13*), and CyberGym (*14*)—designed to assess model-assisted discovery and exploitation of vulnerabilities in open-source codebases. All three share a common evaluation approach: requiring an AI model to iteratively execute tasks related to discovery and exploitation of vulnerabilities within a controlled environment. These evaluation methods do not cover the full risk of AI cyber mis-use as they exclude other relevant attack classes (e.g., cryptanalysis). Nevertheless, this does not mean that they cannot produce suitable evaluations—for a choice of evaluation within one of these three evaluation methods to be suitable, in this case it is arguably enough for it to be sufficiently specific and realistic with respect to a concrete risk scenario, and for there not to be any preliminary evidence indicating a lack of sensitivity or scientific rigor.

## Necessity

HonestCyberEval provides models with direct access to key information, including specific code segments containing synthetically introduced vulnerable lines. This simplification of the discovery and exploitation process diverges from realistic threat environments where such signaling is absent. However, failure remains informative. Concerning its burden, HonestCyberEval is self-contained, uses a simple loop that allows iteration, and keeps setup and per-run costs low.

BountyBench differs from HonestCyberEval in several ways. BountyBench more closely replicates the conditions under which cybersecurity experts operate. It also introduces a wider range of tasks across the attack process, allowing for higher evaluation sensitivity. This helps avoid both benchmark saturation and near-zero performance. Concerning burden, it requires setting up and orchestrating several systems and raises the computational execution costs compared to HonestCyberEval.

CyberGym can evaluate a larger and more diverse set of vulnerabilities compared to HonestCyberEval and BountyBench, further increasing sensitivity and specificity. It achieves highly realistic evaluations of end-to-end discovery and exploitation of vulnerabilities. By covering more instances of the same task, CyberGym also provides higher statistical guarantees, raising its rigor above HonestCyberEval and BountyBench. In terms of burden, CyberGym appears to be the most expensive of the three methods.

Each of these evaluation methods includes evaluations achieving levels of effectiveness that cannot be achieved with less burden by any evaluation from either of the other two methods. In other words, each evaluation method has evaluations that are necessary (see the figure, with Methods 3, 2, and 1 representing HonestCyberEval, BountyBench, and CyberGym, respectively). The shape of the Pareto frontier, which includes components from each of the three methods, depends on the evaluation method but also on the specifics of the model and on the context in which it is developed and deployed.

## Balancing

Each of the three methods has a specific effectiveness–burden trade-off. HonestCyberEval imposes a lower burden and can be understood as a screening evaluation: It simplifies the task with strong model guidance, making failure informative but success only a weak signaling of risk. BountyBench is operationally realistic and allows a gradual assessment of capabilities, enabling evaluators to distinguish modes of failure, but imposes substantial system integration effort. CyberGym can offer high specificity, sensitivity, and rigor through increasing the number of instances per task, which adds to its computational execution cost. These trade-offs need to be considered in the balancing step when choosing which evaluation instance of which evaluation method to require. For example, the regulator might adopt an iterative evaluation approach: requiring an instance of HonestCyberEval unless higher-risk indicators emerge, in which case instances of BountyBench, or even CyberGym, could be required instead to increase confidence in the risk estimate. However, such concrete balancing can only be done in light of all the evidence and context available, which is beyond the scope of this case study. Ultimately, deciding which evaluation provides an appropriate balance between effectiveness and burden requires a scientifically grounded normative judgment.

## RESEARCH CHALLENGES

Applying the principle of proportionality to the requirement for providers to carry out model evaluations for AI risk raises three main research challenges: determine the minimum informational value required for an evaluation to be suitable; advance methodological innovations that can determine which evaluations are "equally effective," enabling the identification of less burdensome evaluations where available; and establish how the resource demands or intrusiveness of evaluations can be lowered without undermining their informational value, and how to approach the unique trade-offs involved for each evaluation. This means that researchers developing AI model evaluations should document and compare the effectiveness and burden of their evaluations to those of other relevant evaluations, to facilitate proportionality assessments.

Amid global concerns that regulatory requirements may hinder innovation and affect competitiveness, the principle of proportionality is highly relevant, ensuring that measures are well calibrated to their objectives and do not impose excessive burden. Its practical application to model evaluations for AI risk requires a scientific approach that enables the regulator to better assess the effectiveness and burden of evaluations and their trade-offs. □

### ACKNOWLEDGMENTS

C.M. and L.M. contributed equally to this work. We thank all participants of the European Commission AI Office Network of Evaluators workshops. The views expressed in this publication do not necessarily reflect those of the European Commission and are made in a personal capacity. The European Commission and any person acting on behalf of it are not responsible for the use that might be made of this publication. V.M. acknowledges funding support from Coefficient Giving. D.L. declares an equity interest in Irregular.

10.1126/science.aea3835

¹European AI Office, European Commission, Brussels, Belgium. ²RAND Corporation, Santa Monica, CA, USA. ³Centre for Health Security, Johns Hopkins University, Baltimore, MD, USA. ⁴Safer AI, Paris, France. ⁵Joint Research Centre, European Commission, Seville, Spain. ⁶CISPA Helmholtz Center for Information Security, Saarbrücken, Germany. ⁷Model Evaluation & Threat Research, Berkeley, CA, USA. ⁸Valencian Research Institute for Artificial Intelligence, Universitat Politècnica de Valencia, València, Spain. ⁹Leverhulme Centre for the Future of Intelligence, University of Cambridge, Cambridge, UK. ¹⁰Irregular, San Francisco, CA, USA. ¹¹Italy's National Cybersecurity Agency, Rome, Italy. ¹²The Alan Turing Institute, London, UK. ¹³Apart Research, San Francisco, CA, USA. Email: carlos.mougan@ec.europa.eu

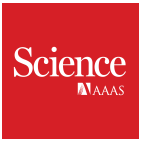

# The science and practice of proportionality in AI risk evaluations

Carlos Mougan, Lauritz Morlock, Jair Aguirre, James R. M. Black, Jan Brauner, Simeon Campos, Sunishchal Dev, David Fernández Llorca, Alberto Franzin, Mario Fritz, Emilia Gómez, Friederike Grosse-Holz, Eloise Hamilton, Max Hasin, Jose Hernandez-Orallo, Dan Lahav, Luca Massarelli, Vasilios Mavroudis, Malcolm Murray, Patricia Paskov, Jaime Raldua, and Wout Schellaert

**View the article online**
https://www.science.org/doi/10.1126/science.aea3835
**Permissions**
https://www.science.org/help/reprints-and-permissions